\documentclass[prl,reprint,superscriptaddress,showpacs,preprintnumbers,nobibnotes,amsmath,amssymb, aps,longbibliography]{revtex4-2}

\usepackage{cancel}
\usepackage{accents}
\usepackage{mciteplus,slashed}
\usepackage{amssymb,cancel,amsmath}
\usepackage{dcolumn}
\usepackage{bm}
\usepackage[caption=false]{subfig}
\usepackage{appendix}
\usepackage{feynmp-auto}
\usepackage[T1]{fontenc}	
\usepackage{csvsimple}
\usepackage[colorlinks=true,citecolor=blue,linkcolor=blue]{hyperref}
\usepackage[section]{placeins}
\usepackage[capitalise]{cleveref}
\usepackage{booktabs}
\usepackage{graphicx}
\usepackage{subfig}
\usepackage{mathrsfs}
\usepackage{multirow}
\usepackage[utf8]{inputenc}
\usepackage{siunitx}
\usepackage{comment}
\usepackage{gensymb}
\usepackage{enumitem}

\usepackage[dvipsnames]{xcolor}
\usepackage[normalem]{ulem}

\begin{document}

\title{Detecting High-Energy Neutrinos from Galactic Supernovae with ATLAS}

\begin{abstract}
We show that ATLAS, a collider detector, can measure the flux of high-energy supernova neutrinos, which can be produced from days to months after the explosion. 
Using Monte Carlo simulations for predicted fluxes, we find at most $\mathcal{O}(0.1-1)$ starting events and $\mathcal{O}(10-100)$ throughgoing events from a supernova 10 kpc away. 
Possible Galactic supernovae from Betelgeuse and Eta Carinae are further analyzed as demonstrative examples. 
We argue that even with limited statistics, ATLAS has the ability to discriminate among flavors and between neutrinos and antineutrinos, making it an unique neutrino observatory so far unmatched in this capability.
\end{abstract}

\author{Alex~Y.~Wen}
\email{alexwen@g.harvard.edu}
\affiliation{Department of Physics \& Laboratory for Particle Physics and Cosmology, Harvard University, Cambridge, MA 02138, USA}

\author{Carlos~A.~Arg{\"u}elles}
\email{carguelles@fas.harvard.edu}
\affiliation{Department of Physics \& Laboratory for Particle Physics and Cosmology, Harvard University, Cambridge, MA 02138, USA}

\author{Ali~Kheirandish}
\email{ali.kheirandish@unlv.edu}
\affiliation{Department of Physics \& Astronomy, University of Nevada, Las Vegas, NV 89154, USA}
\affiliation{Nevada Center for Astrophysics, University of Nevada, Las Vegas, NV 89154, USA}

\author{Kohta~Murase}
\email{murase@psu.edu}
\affiliation{Department of Physics, Department of Astronomy \& Astrophysics, \& Center for Multimessenger Astrophysics, Institute for Gravitation and the Cosmos, The Pennsylvania State University, University Park, PA 16802, USA}
\affiliation{School of Natural Sciences, Institute for Advanced Study, Princeton, NJ 08540, USA}
\affiliation{Center for Gravitational Physics and Quantum Information, Yukawa Institute for Theoretical Physics, Kyoto, Kyoto 16802, Japan}

\maketitle

\section{Introduction}

The discovery of high-energy astrophysical neutrinos, first reported by IceCube in 2013~\cite{IceCube:2013cdw,IceCube:2013low}, opened a new window to the Universe and marked the start of an era of multimessenger astrophysics.
Cosmic neutrinos are valuable probes of astrophysical processes~\cite{Ackermann:2019ows,DeGouvea:2020ang} and neutrino physics~\cite{Arguelles:2019rbn,Ackermann:2022rqc,deGouvea:2019goq}.
However, small neutrino cross sections~\cite{Formaggio:2012cpf} and the observed falling energy spectra~\cite{IceCube:2020wum} have so far limited their study to very large volume detectors proposed or built in naturally occurring media such as glaciers~\cite{ANITA:2016vrp,ANITA:2018sgj,IceCube-Gen2:2020qha}, lakes~\cite{BAIKAL:2015hjt,Avrorin:2015wba}, oceans~\cite{KM3Net:2016zxf,P-ONE:2020ljt,Ye:2022vbk}, or mountains~\cite{Sasaki:2017zwd,Brown:2021lef,Thompson:2023pnl}.
These detectors are sparser, and have relatively poor energy and angular resolution, and particle identification capabilities, compared to more densely instrumented detectors used in collider physics.

Even with limited statistics, IceCube measurements of the astrophysical neutrino flavor composition have already yielded some of the strongest constraints on long-range forces~\cite{Bustamante:2018mzu}, quantum-gravity operators~\cite{Arguelles:2015dca,IceCube:2021tdn}, the neutrino lifetime~\cite{Shoemaker:2015qul,Bustamante:2016ciw,Song:2020nfh,Abdullahi:2020rge}, and ultralight dark matter interactions~\cite{Farzan:2018pnk,Reynoso:2022vrn,Arguelles:2023wvf}, to name a few of many models~\cite{Arguelles:2019tum,Carloni:2022cqz,Arguelles:2022tki}.  

Further information can be obtained if astrophysical neutrinos are detected by collider detectors, and transient neutrino sources may provide unique opportunities~\cite{Murase:2019xqi,Murase:2019tjj,Guepin:2022qpl}.
In particular, the next Galactic supernova (SN) has been expected to yield a large detectable neutrino signal in the GeV--TeV range, and neutrino detection with large statistics at multienergies is possible~\cite{Murase:2017pfe}. 

In this \textit{Letter}, we show that large collider detectors serve as unique astrophysical neutrino telescopes, which enables, among other things, precise measurements of the flavor ratio of astrophysical neutrinos. To demonstrate this, we consider ATLAS ~\cite{ATLAS:2008xda,ATLAS:2023dns}, a barrel-shaped multi-purpose detector situated at the Large Hadron Collider (LHC) at CERN, primarily designed to study reactions originating at a beam collision point.
ATLAS possesses a sensitive, massive hadronic calorimeter~\cite{ATLAS:1996aa,ATLAS:2018edp,ATLAS:2008xda,ATLAS:2020cli} making it a viable fiducial volume for energetic neutrino events, and a sophisticated muon spectrometer~\cite{ATLAS:2008xda,ATLAS:2016lqx,ATLAS:2020auj,ATLAS:2010xrj,ATLAS:1997ad} surrounding the calorimeter, capable of identifying muon tracks and measuring their momenta.
This detector combination makes neutrino detection viable.


\section{High-Energy Neutrino Emission from Supernovae} \label{section_SN}

Neutrinos play a critical role in the dynamics of a SN explosion. In addition to the known~\cite{Janka:2017vlw,Bethe:1990mw,Woosley:2002zz} and detected~\cite{Kamiokande-II:1987idp,Bionta:1987qt} prompt flux of MeV neutrinos, core-collapse SNe are also promising sources of high-energy neutrinos~\cite{Murase:2017pfe,Murase:2023chr}.
Recent SN observations, especially in the optical band, provided strong evidence that interaction with dense, confined circumstellar material (CSM) transiently occurs as the SN shock wave propagates outwards~\cite{Yaron:2017umb,Morozova:2017hbk,Hosseinzadeh:2017uig,Boian2020, Tinyanont:2021cwl,Bostroem:2023dvn}.
Older SN remnants (with ages of $10^2-10^3$~yr) have been established as cosmic-ray accelerators~\cite{Funk:2015ena,Caprioli:2023orv}, and interacting SNe may also efficiently emit high-energy neutrinos and gamma rays~\cite{Murase:2010cu,Murase:2023chr}. 
For the next Galactic SN, even ordinary SNe like Type II-P SNe would produce sufficiently large fluxes of neutrinos that are detectable to many terrestrial neutrino detectors such as IceCube~\cite{Murase:2017pfe}, and even minibursts from nearby galaxies could be observed~\cite{Kheirandish:2022eox,Valtonen-Mattila:2022nej}. 
The time window of neutrino signals is predicted to be the 10--100 day time range following an explosion~\cite{Murase:2017pfe,Murase:2023chr}. 

Type II-P and IIn SN make up approximately 50\% and 3-7\%, respectively, of all core-collapse SNe~\cite{Graur:2016lca}.
The result for SNe II-P may also hold for other SN types (II-L, IIb) as long as they have a sufficiently dense CSM~\cite{Margutti:2016wyh,Murase:2018okz}, so studying these two SN types would be representative of most SNe with confined CSM.

Predicted neutrino fluxes from SNe have uncertainties which primarily depend on CSM properties. The CSM density is written as $\rho_{\rm cs}=D r^{-2}$ for a wind-like density profile, where $D \equiv 5\times 10^{16}\: \text{g}\:\text{cm}^{-1} D_*$ is the CSM parameter and $r$ is the radius from the center of the SN explosion. We consider the range of $0.01 < D_* < 1.0$ for SN II-P, and $0.1 < D_* < 1.0$ for SN IIn. This is sufficient for the purpose of this work to demonstrate the feasibility of ATLAS-like detectors for detecting astrophysical neutrinos, and other parameters, such as the spectral index, only moderately affect the overall detectability or have degeneracies with $D_*$. See Refs.~\cite{Murase:2017pfe,Kheirandish:2022eox} for details.   

The studied values of $D_*$ are determined by SNe observations, which suggest a range of $D_*\sim0.01-1$. For example, one of the canonical examples is SN 2013fs, which has $D_*\sim0.01$ \cite{Boian2020}. 
The recent event SN 2023ixf has $D_*\sim0.1$ \cite{Jacobson-Galan:2023ohh}, while SN 2020tlf has $D_*\sim1$ \cite{Jacobson-Galan:2021pki}. 

\section{High-Energy Neutrino Events in ATLAS} \label{section_methods}
High-energy neutrinos may interact within the detector itself (\textit{starting} events), or produce a muon originating from an interaction within the Earth (\textit{throughgoing} events). 
For starting events, a charged-current (CC) or neutral-current (NC) deep inelastic scattering (DIS) would leave an energetic hadronic recoil within the ATLAS hadronic calorimeter.
An accompanying muon may also be detected by the ATLAS muon spectrometer.
For thoughgoing events, signals can only come from $\nu_\mu$ CC interactions in surrounding bedrock, with a subdominant contribution from $\nu_\tau$ (for simplicity, however, the $\tau$ component is ignored); ATLAS may detect these muons as they traverse the muon spectrometer. 


\begin{figure*}[t!]%
    \centering
    \subfloat[\centering starting events]{\includegraphics[width=0.49\linewidth]{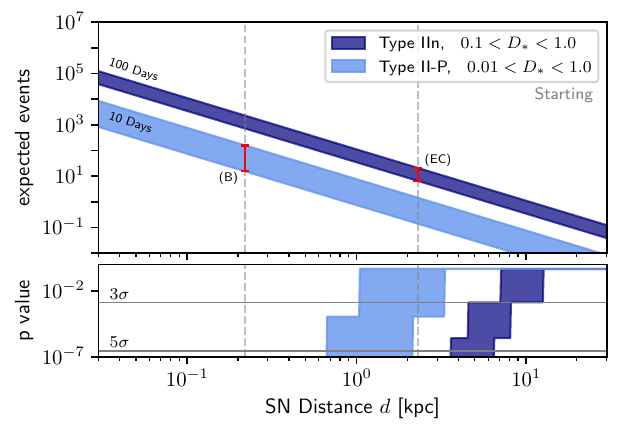}}%
    \quad
    \subfloat[\centering throughgoing events]{\includegraphics[width=0.49\linewidth]{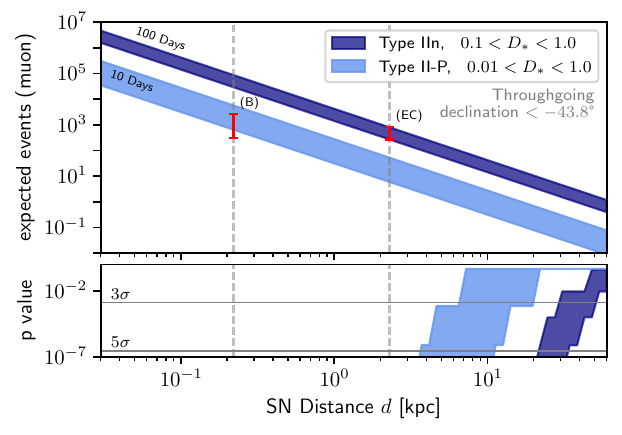}}%
    \caption{\textbf{\textit{Event rates and observation significance of high-energy supernovae neutrinos in ATLAS.}} The throughgoing events rates represent the maximum number for a source that is always below the horizon. Below each panel, a plot of p-values for rejecting the atmospheric neutrino background-only hypothesis is shown. The estimated rates for Betelgeuse-like (B) and Eta Carinae-like (EC) SNe scenarios are shown in red bars.}%
    \label{fig:distance}%
\end{figure*}

The expected number $\mathcal{N}$ of starting events induced by neutrinos, in a volume of mass $m$, is given by
\begin{equation}\label{eq:expected_events}
{\mathcal N} = \int dE_\nu \int dt \: \dot{\phi}_{\nu} (E_\nu,t)  \: \sigma_{\nu-\rm nuc}(E_\nu) N_{\text{nuc}}(m),
\end{equation}
where $\dot{\phi}_\nu(E_\nu,t) = (dN_\nu / dE_\nu dt)/(4\pi d^2)$ is the all flavor neutrino flux, $dN_\nu / dE_\nu dt$ depends on models (e.g., via $D_*$), $d$ is the distance to a SN, $\sigma_{\nu-\rm nuc}(E_\nu)$ is the neutrino-nucleon cross section, and $N_{\rm nuc}(m)$ is the number of nucleon targets in the fiducial volume.

For high-energy neutrinos, DIS dominates the total cross section $\sigma_{\nu-\rm nuc}$, and we assume that matter is made of iso-scalar targets; cross section is averaged over the neutrino-proton and neutrino-neutron values. 
Neutrinos and antineutrinos are computed separately owing to their distinct cross sections.
The integral in Eq. \ref{eq:expected_events} is taken over the energy range $[10^2,\: 10^6]\:\si\GeV$.
We expect the detection of starting events to be analogous to existing ATLAS studies~\cite{ATLAS:2021kxv,PompaPacchi:2021yjp} that utilize the missing transverse energy trigger \cite{ATLAS:2020atr}, which is only most efficient above $200 \:\si\GeV$~\cite{ATLAS:2021kxv,PompaPacchi:2021yjp}. 
Our energy range is chosen to reflect that, since we expect similar triggering for neutrino events.
At high-energy, $E_\nu \dot{\phi}_\nu$ approximately falls with $E_\nu^{-1}$, which yields a negligibly small rate above $10^6~\si\GeV$.

When integrating in time, we conservatively take 100 days for SNe IIn and 10 days for SNe II-P, based on the signal-to-background calculation in Ref.~\cite{Murase:2017pfe} as indicative of the characteristic time windows to search for neutrino signals. 
For ATLAS, we assume the hadronic calorimeter mass $m=4000$ metric tons, and include both CC and NC contributions in $\sigma_{\nu-\rm nuc}(E_\nu)$ when computing the number of starting events. 

Throughgoing events are estimated with a Monte Carlo method using techniques described in Ref.~\cite{IceCube:2020tcq}.
Using \texttt{LeptonInjector}~\cite{IceCube:2020tcq}, we generate a large quantity of $\nu_\mu$ CC interactions in a 10 km long rock column preceding the detector.
Generated muons are propagated through rock to the detector using \texttt{PROPOSAL}~\cite{Koehne:2013gpa}. 
The total number of expected interactions in the rock is calculated with ~\cref{eq:expected_events} and scaled by the fraction of muons that propagated to the detector, to obtain a physical throughgoing event rate \footnote{For more details, see Supplemental Methods and Tables which include Refs. \cite{ATLAS:2020auj,ATLAS:2011tau,Gonzalez-Garcia:2013iha,Halzen:2016seh,Kheirandish:2022eox,ParticleDataGroup:2020ssz,ATLAS:2011tau,Gonzalez-Garcia:2013iha,Halzen:2016seh,Kheirandish:2022eox,IceCube:2020tcq,Koehne:2013gpa,Vincent:2017svp,NuFluxRepository,Fedynitch:2022vty,Gaisser:2013bla,Riehn:2017mfm,Kopp:2007ai}}. 

Efficiently detecting throughgoing muons will require novel trigger development. It should be sensitive to the directionality of the incoming muons (below the horizon) and also to the fact that muons would enter and exit the barrel-shaped spectrometer while not necessarily traversing the beam collision point.


The dominant background consists of atmospheric neutrinos.
We estimate the starting and throughgoing background separately with the same method described above, but with atmospheric fluxes instead. 
These fluxes are further described in the Supplemental Methods.

We anticipate the LHC beam to not constitute a significant background, even if the beam is running. 
The segmentation of the hadronic calorimeter and position information from the muon spectrometer can determine the directionality of signal events, and distinguish them from possible beam-induced backgrounds originating from the collision point.
Starting events may see potential hadronic backgrounds induced by cosmic muons \cite{ATLAS:2016ndt}; we also expect directionality correlation with a SN source to mitigate this.

\section{Results} \label{section_results}
We evaluate the event rates and the significance of observing high-energy neutrinos from two representative types of core-collapse SNe (IIn and II-P) in ATLAS.
The expected numbers of signal events varying with distance are shown for starting and throughgoing events in the top panels of~\cref{fig:distance}; in the bottom panels, we also show the p-values for rejecting the background-only hypothesis. 

Starting events for SNe II-P (IIn) would only constitute a significant signal if the SN was closer than approximately 0.6~kpc (3~kpc) (small compared to the $\sim25$~kpc size of our Galaxy).
However, throughgoing events are produced by a larger effective volume of the target, provided that the source is below the horizon for a sufficient period of time for neutrinos to interact in the bedrock around the detector.
Optimistically, for a source that is always below the horizon, throughgoing events enable the detection horizon for SNe II-P (IIn) up to around 4~kpc (20~kpc). 
Cases of two close-by stars Betelgeuse~\cite{Dolan:2016,Smith:2008ef,Meynet:2013pqa} and Eta Carinae \cite{Smith:2008gx,Smith:2012nw} as prospective Type II-P and IIn SNe candidates and shown as demonstrative examples of interest. 

\begin{figure}
\centering
\includegraphics[width=\linewidth]{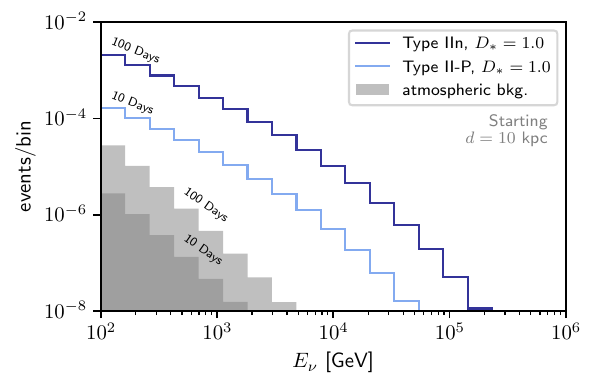}\\
\includegraphics[width=\linewidth]{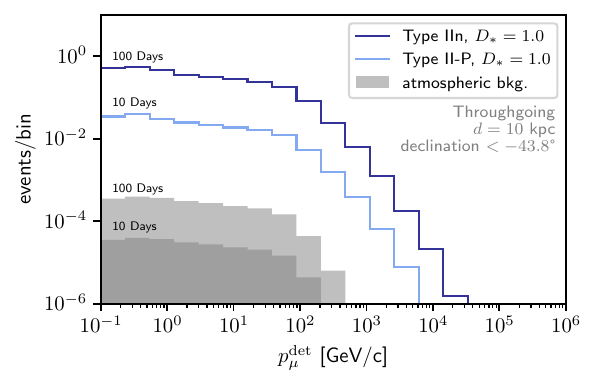}
\caption{\textbf{\textit{Throughgoing and starting event energy distributions.}} 
Top: Neutrino energy $E_\nu$ distribution of starting events in ATLAS for Type IIn and II-P SN with $D_*=1$ at distance $d=10\:\text{kpc}$ for 100 (dark blue) and 10 days (light blue) of data taking.
Bottom: Distribution of the muon momentum, $p_\mu^{\rm det}$, at the detector for throughgoing events.
The shape of the spectrum is due to the consideration of \SI{100}{\GeV} neutrino events and above, which produces a flatter distribution of muons at lower energies.
For both plots, corresponding background from atmospheric neutrinos are shown as shaded gray regions.
}
\label{fig:edist}
\end{figure}

Neutrino energy distributions for starting events at the interaction point are shown in the top panel of~\cref{fig:edist}.
These spectra adopt the same shape for both starting and throughgoing events, although they would not be measurable for throughgoing events due to muon energy losses.
The estimated atmospheric neutrino background is also shown in the same figure, integrated over both 10 and 100 days to directly compare with the corresponding SN cases.

We show the muon momentum $p_\mu^{\textrm{det}}$ spectrum of throughgoing muons at the detector in the bottom panel of~\cref{fig:edist}. 
The relation between this spectrum to the commonly-measured transverse momentum $p_T$ will depend on the orientation of the detector relative to the direction of the incoming neutrino flux; we assume that the flux arrives sideways on (perpendicular to the beam axis). 
The momentum spectrum also gives an idea of the distribution of muon \textit{sagitta} \cite{ATLAS:1997ad} that should be expected in the magnetized part of the detector. 

A key characteristic of this signal is the directionality of the muons, from below the horizon; this would not be produced by cosmic muon backgrounds, and only a small background is produced by atmospheric muon neutrinos.
This background is also shown in~\cref{fig:edist} (bottom).
With an assumed throughgoing angular resolution of $5\degree$, a signal should be well-correlated to a SN point source.

\begin{figure*}
\centering
\includegraphics[width=0.8\paperwidth]{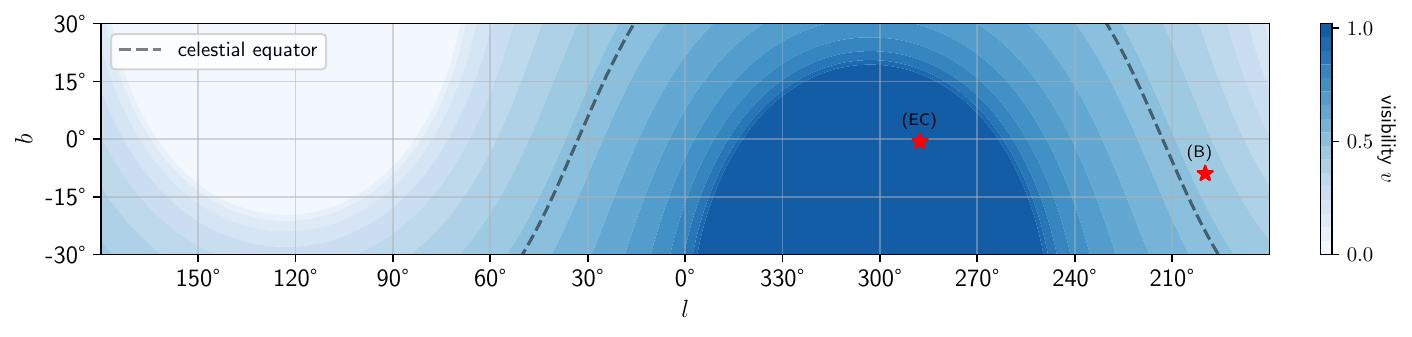}
\caption{\textbf{\textit{Throughgoing events visibility for ATLAS.}} 
Skymap, in Galactic coordinates, showing the visibility, $v$, of throughgoing events depending on the position in the sky.
The locations of Betelgeuse (B) and Eta Carinae (EC) are marked by red stars.
The dashed line indicates the celestial equator.
}
\label{fig:atlas-skymap}
\end{figure*}

For the throughgoing events presented in both~\cref{fig:distance}(b) and \ref{fig:edist} (bottom), we have assumed a SN source that is always below the horizon over the course of the 10 or 100~day observation period. 
However, this will not be the case for every SN event, given that the ATLAS detector, at a latitude of $46.2\degree$, will only see throughgoing events $100 \%$ the time from objects in the celestial sky with a declination of $\delta<-43.8\degree$. 
We define the visibility factor, $v$, of a celestial coordinate to be the fraction of time that it is below the horizon at the ATLAS latitude; hence any object with $\delta<-43.8 \degree$ will have $v=1$. 
The value of $v$ will decrease until $\delta>43.8 \degree$, where $v=0$.
\Cref{fig:atlas-skymap} shows the value of $v$ in galactic coordinates; in order to determine an event estimate for throughgoing events, the event number must be scaled by $v$. 

We also consider specific cases to illustrate a more concrete scenario of a hypothetical SN explosion: a Betelgeuse-like (B), and a Eta Carinae-like (EC), SN explosion which occur at distances of 0.22 kpc and 2.3 kpc respectively.
For (B) and (EC), we use $0.01 < D_* < 1.0$ (assuming a SN II-P) and $0.1 < D_* < 1.0$ (assuming a SN IIn), respectively. 

The results from these hypothetical signals are indicated in~\cref{fig:distance}: for (B) we anticipate 15-150 (300-2,600) starting (throughgoing) events, and for (EC) we anticipate 6-21 (170-800) starting (throughgoing) events. The throughgoing signal for (B) is multiplied by a visibility factor of $v=0.46$ due to its location in the sky. The celestial positions of (B) and (EC) are shown in~\cref{fig:atlas-skymap}, mapped to a corresponding visibility $v$. 

\section{Discussion and Conclusion} \label{section_conclusions}

In addition to the demonstrated feasibility of ATLAS as a unique detector for astrophysical high-energy neutrinos, we also anticipate comparable capabilities for similar detectors like CMS~\cite{CMS:2008xjf}. Any kiloton-scale or larger, densely-instrumented, present or future detector may consider the prospect of detecting high-energy neutrinos from Galactic SNe. 

As a previous effort to characterize ATLAS as a viable detector of natural neutrinos, Ref.~\cite{Kopp:2007ai} studied the precision measurement of atmospheric neutrinos at a lower energy. Although the expected sample size was small, it highlights the advantages of using a precision collider detector for neutrino physics. 

\begin{figure}[htb!]
\centering
\includegraphics[width=0.85\linewidth]{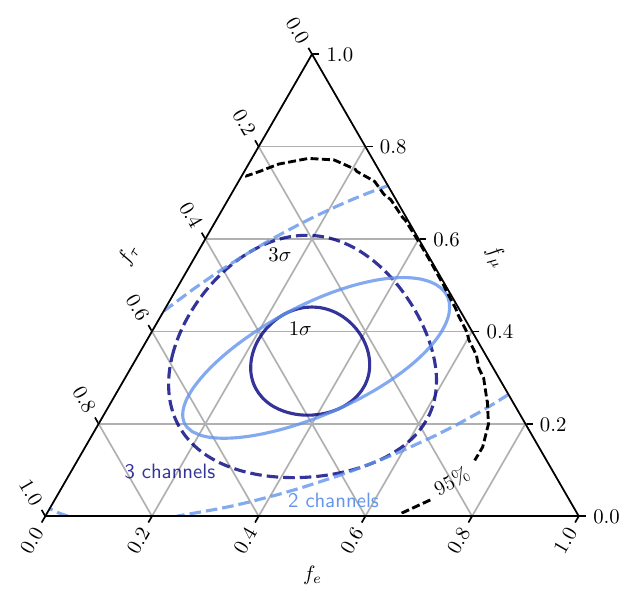}
\caption{\textbf{\textit{Expected flavor triangle allowed region by ATLAS.}}
The flavor triangle measured at Earth by ATLAS using only starting events from a single close by SN explosion, similar to Betelgeuse (B).
The dark blue lines correspond to performing an analysis using all 3 signal channels described in the text, whereas the light blue lines only use channels (1) and (2).
Solid and dashed lines indicate $1\sigma$ and $3\sigma$ confidence, respectively.
The dashed black line shows the 95\% confidence interval from IceCube using the HESE sample, taken from Fig. 6 of Ref. \cite{IceCube:2020fpi}.}
\label{fig:flavor_triangle_starting}
\end{figure}

Given ATLAS' unique instrumentation often unseen in dedicated neutrino detectors, it may be possible to discriminate between all three neutrino flavors. 
Consider a benchmark scenario with 88 (22 NC and 66 CC) starting events, which is roughly the expected signal from (B) with $D_*=0.1$.
We can broadly consider three distinguishable signal channels: (1) one hadronic shower, (2) one hadronic shower plus muon, and (3) two hadronic showers. 
Each flavor of starting events will contribute to these channels, allowing us to estimate the expected signal in each channel and infer the flavor ratio $(f_e,f_\mu,f_\tau)$.
In~\cref{fig:flavor_triangle_starting} we show the allowed flavor ratios when assuming a $(1,1,1)$ ratio flavor composition at Earth; we also show a more pessimistic 2-channel case assuming no sensitivity to channel (3) events.
A better understanding of the detector efficiency for throughgoing muons is required to incorporate throughgoing signal (muon-flavor only) into this measurement.
An ATLAS flavor ratio measurement is expected to be comparable to, or better than, current measurements by dedicated experiments~\cite{Song:2020nfh,IceCube:2020fpi}; ~\cref{fig:flavor_triangle_starting} also shows the 95\% confidence interval of an IceCube flavor measurement \cite{IceCube:2020fpi} using the HESE sample consisting of high energy all-sky astrophysical starting events \cite{IceCube:2020wum}.
Future large-scale experiments like Hyper-Kamiokande \cite{Hyper-Kamiokande:2018ofw} with significantly more statistics and sophisticated event topologies may offer better constraints.

Another advantage of ATLAS is a superior energy resolution compared to that of dedicated neutrino detectors.
The ATLAS hadronic calorimeter energy resolution for jets is approximately given by $\sigma/E = 50\% / \sqrt{E/\text{GeV}} \oplus 3\%$~\cite{ATLAS:2018edp}, translating to approximately $1.6\% \oplus 3\%$ at \SI{1}{\TeV}.
This can be contrasted to, for example, the IceCube energy resolution of $\sim$$15\%$ for shower events~\cite{IceCube:2013dkx}, around an order of magnitude worse.

Finally, ATLAS is expected to have the capability for neutrino-to-antineutrino separation.
Assuming a typical path length of around \SI{5}{\meter} through the ATLAS muon spectrometer barrel, which is magnetized at approximately \SI{0.5}{\tesla}, a \SI{1}{\TeV} muon track will have a sagitta of approximately \SI{500}{\micro\meter}~\cite{2106380}, well above the $\sim$30-\SI{40}{\micro\meter} spectrometer alignment accuracy and the $\sim$80-\SI{90}{\micro\meter} detector single hit resolution quoted in Refs.~\cite{ATLAS:2016lqx,2106380}. 
Only at approximately $\SI{5}\TeV$ will the sagitta approach $\sim$\SI{100}{\micro\meter}, a length scale limited by the detector and alignment resolutions. 
Since the bulk of muons from both starting and throughgoing events are expected to be less energetic, ATLAS hardware can likely determine the charge of most muons that traverse it.
If successful, ATLAS may yield a unique event-by-event measurement of the ratio of neutrinos to antineutrinos, which can be used to discriminate between different production mechanisms at the source.
While larger experiments like IceCube or Hyper-Kamiokande may perform a statistical measurement of this ratio (difficult due to the difficulty of measuring inelasticity at high energies), they lack magnets for charge discrimination on an event-by-event basis.

In conclusion, the event rates and estimated hardware capabilities of ATLAS make it a promising high-energy neutrino telescope. 
We hope that our findings spur the development of new triggers and analyses to enable a precise measurement of the next nearby SN event.

\section*{Acknowledgements}

We thank Austin Schneider and Nicholas Kamp for their help with \texttt{LeptonInjector}. 
We thank Masahiro Morii and Melissa Franklin for insightful discussions about the capabilities of ATLAS.
We thank the KITP for being an engaging space to work on physics, and this research was supported in part by the National Science Foundation under Grants No. NSF PHY-1748958 and PHY-2309135.
CAA is supported by the Faculty of Arts and Sciences of Harvard University and the Alfred P. Sloan Foundation.
AK is supported by the NASA Grant 80NSSC23M0104.
AYW is supported by the Harvard Physics Department Purcell Fellowship and the Natural Sciences and Engineering Research Council of Canada (NSERC), funding reference number PGSD-577971-2023.
KM is supported by the NSF Grant No.~AST-2108466 and No.~AST-2108467, and KAKENHI No.~20H01901 and No.~20H05852.

\clearpage
\bibliography{main}
\clearpage




\newpage

\onecolumngrid
\appendix

\ifx \standalonesupplemental\undefined
\setcounter{page}{1}
\setcounter{figure}{0}
\setcounter{table}{0}
\setcounter{equation}{0}
\fi

\renewcommand{\thepage}{Supplemental Methods and Tables -- S\arabic{page}}
\renewcommand{\figurename}{SUPPL. FIG.}
\renewcommand{\tablename}{SUPPL. TABLE}

\renewcommand{\theequation}{A\arabic{equation}}

\section{Supplemental Methods and Tables}

\subsection{Sensitivity to flavor ratio}

While the hadronic calorimeter has demonstrated ability to detect lepton tracks~\cite{ATLAS:2020auj}, the energy deposit per length is still small, and is relatively unstudied at energies considered in this work; hence to be conservative, we assume the hadronic calorimeter is only sensitive to hadronic showers caused by nuclear recoils from CC or NC interactions, with a perfect efficiency. 
We assume this efficiency is one, or at least close to one, since we do not require sophisticated reconstruction of particles, but only a large energy deposit in a localized region. 

For $\mu$-flavor events, we will assume an overall muon spectrometer efficiency of $\varepsilon_{\mu}=0.75$, approximately the reconstruction and selection efficiency at 0.5-\SI{1}{\TeV} for high-$p_T$ working point muons reported in Ref.~\cite{ATLAS:2020auj}.
The detection of muons is expected to rely only on the muon spectrometer, and not the inner detector which is also used to reconstruct tracks during normal detector operation.

Finally, for $\tau$-flavor events we assume that we can only detect a hadronic $\tau$ decay if it propagates $\SI{0.5}\m$ or less - corresponding to an energy of less than $\sim 10^4\si\GeV$ which includes around $\varepsilon_{\tau\:\text{range}}=90\%$ of all $\tau$ events produced in all SN scenarios considered. 
If the $\tau$ decays into a muon or electron, we assume the signal is indistinguishable from a $e$ or $\mu$-flavor event.

Thus we identify three possible signal channels: (1) only a hadronic shower, from NC, $\nu_e$, and $\tau\rightarrow e$ events; (2) a hadronic shower and a muon, from mostly $\nu_\mu$ and $\tau \rightarrow \mu$ events; and (3) two hadronic showers separated by some distance, caused by hadronic $\tau$ events. These signal channels are summarized in Suppl. Table~\ref{tab:channels}.
Each flavor of starting event will contribute to these channels in varying amounts, taking into account the estimated detector efficiencies. 
This allows us to infer the flavor ratio.

\begin{table}[]
\caption{\label{tab:channels}A summary of the three starting event signal channels that may be used for flavor measurements. $N_{NC}$ NC and $N_{CC}$ CC events, required for estimating the event number in each channel, are estimated using the methods described in the text for starting events.}
\begin{tabular}{cccc}
\hline
Channel                             & 1                 & 2                      & 3                                 \\ \hline
Signal                              & Hadronic shower   & Hadronic shower + muon & Hadronic shower + Hadronic shower \\ \hline
\multirow{3}{*}{Physical processes} & All NC events     &  $\nu_\mu$ CC events      &  $\nu_\tau$ CC events + $\tau$ hadronic decay \\
                                    & $\nu_e$ CC events   & $\nu_\tau$ CC events + $\tau\rightarrow \mu$ decay&                                   \\
                                    & $\nu_\tau$ CC events + $\tau\rightarrow e$ decay &                        &                                   \\ \hline
\end{tabular}
\end{table}


To estimate the discovery p-value of rejecting a $(f_e,f_\mu,f_\tau) = (1,1,1)$ null hypothesis, we employ a similar test used in Refs.~\cite{ATLAS:2011tau,Gonzalez-Garcia:2013iha,Halzen:2016seh,Kheirandish:2022eox}. 
We define the test statistic
\begin{equation}\label{eq:q0def}
q_0 = -2 \log \mathcal{L} = 2\sum_i \left( Y_i - N_i + N_i\log \left( \frac{N_i}{Y_i} \right) \right),
\end{equation}
where $i$ runs over each channel as summarized in Suppl. Table~\ref{tab:channels}. $Y_i$ and $N_i$ are the expected number of events given a $(1,1,1)$ and varying $(f_e,f_\mu,f_\tau)$ ratio, respectively.
The test statistic value corresponding to $1\sigma$, $3\sigma$ p-values is looked up in the table for the joint estimation of two parameters in chapter 40 of the Review of Particle Physics~\cite{ParticleDataGroup:2020ssz}.


\subsection{Calculation of discovery p-value}

The discovery p-value of rejecting the background hypothesis for starting and throughgoing events presented in~\cref{fig:distance} is calculated in the same way as in Refs.~\cite{ATLAS:2011tau,Gonzalez-Garcia:2013iha,Halzen:2016seh,Kheirandish:2022eox}. 
The test statistic is again given by~\cref{eq:q0def} where $Y_i$ and $N_i$ are the expected number of events calculated with the background only hypothesis and background-plus-signal hypothesis, respectively.
$i$ only runs over one bin which contains all the events.
The p-value $p$ is given by 
\begin{equation}\label{eq:p-value-def}
p = \frac{1}{2} \left( 1 - \text{erf}\left( \sqrt{\frac{q_0}{2}} \right)  \right).
\end{equation}

\subsection{Simulation of throughgoing events}

\begin{figure*}[htb!]
\centering
\includegraphics[width=0.4\paperwidth]{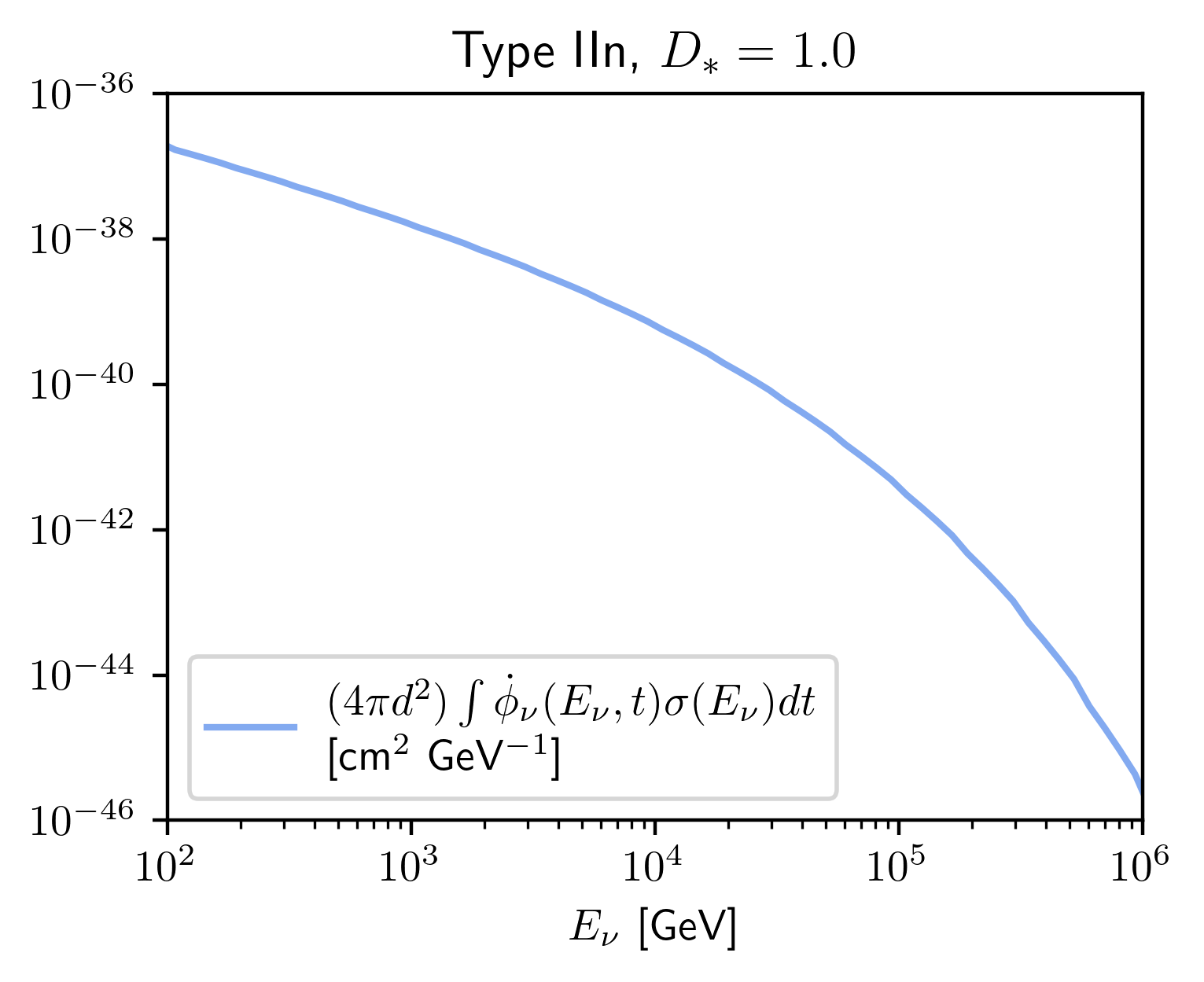}
\caption{\textbf{\textit{Example (and un-normalized) energy distribution used for event generation.}} 
$(4\pi d^2)\int \dot{\phi}_\nu(E_\nu,t)\sigma_{\nu-\rm nuc}(E_\nu) dt$ with $\dot{\phi}_\nu$ corresponding to a SN IIn, $D_*=1.0$ SN scenario. 
}
\label{fig:energy_density}
\end{figure*}

\begin{figure*}[htb!]
\centering
\includegraphics[width=0.83\paperwidth]{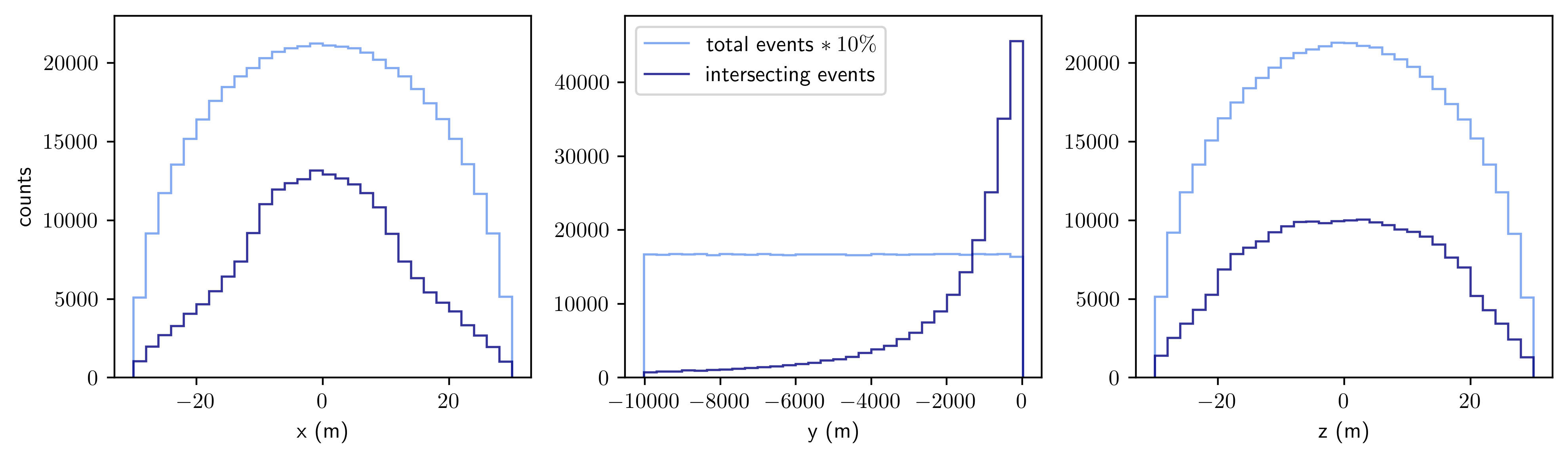}
\caption{\textbf{\textit{Positions of vertices generated using \texttt{LeptonInjector}.}} 
The neutrino CC interaction vertices plotted in each spatial coordinate, uniformly distributed in a cylinder of radius \SI{30}{\meter} and length \SI{10}{\kilo \meter}; this cylinder lies along the y-axis. ATLAS is treated as a barrel-shaped detector with the beam axis along the z-axis. The light blue lines correspond to all $N'_{\rm sim}$ vertices generated (counts scaled by $0.1$) and the dark blue lines correspond only to vertices with a muon that intersects the detector. 
}
\label{fig:vertex_distribution}
\end{figure*}

Throughgoing events are estimated with a Monte Carlo method using techniques described in Ref.~\cite{IceCube:2020tcq}.
First, a large number $N'_{\rm sim}$ of neutrino interaction vertices is generated uniformly in a cylindrical rock column with density $\rho_{rock}=2.65\:\text{g}\:\text{cm}^{-3}$ resembling earth's crust, of dimensions 10 km length and 30 m radius with ATLAS at one end.
Muons are thus exposed to a detector footprint of dimensions $\text{diameter}\times\text{length}=22 \: \text{m} \times 40 \:\text{m}$. 
In Suppl.~Fig.~\ref{fig:vertex_distribution}, we show the distribution of all such generated vertices, and of the vertices corresponding to a produced muon that intersects the detector. 
These vertices are generated according to a neutrino energy distribution $(4\pi d^2)\int \dot{\phi}_\nu(E_\nu,t)\sigma_{\nu-\rm nuc}(E_\nu) dt = \int \frac{dN_\nu}{dE_\nu dt}(E_\nu,t)\sigma_{\nu-\rm nuc}(E_\nu) dt$, with the appropriate $\phi_\nu$ and integration time for each SN scenario.
The cross section is only for CC muon flavor interactions.
An plot of this density is shown in Suppl.~Fig.~\ref{fig:energy_density}, for the example case of a Type IIn, $D_*=1.0$, SN.

This event generation process is done using \texttt{LeptonInjector}~\cite{IceCube:2020tcq}, which, after vertex-generation, samples the energy of the out-going muon from differential cross sections, and records only the muons which point in a direction that intersects the detector volume.
Muons that intersect the detector volume are then propagated using \texttt{PROPOSAL}~\cite{Koehne:2013gpa} until they reach a low-energy threshold on the scale of the muon mass.
\texttt{PROPOSAL} considers both stochastic and continuous muon energy losses. 
Muons energetic enough to reach the detector are recorded and are considered part of the throughgoing signal, of which there are $n$ events.

The physical event number is estimated by calcuating the total number of interaction vertices expected in the rock using \cref{eq:expected_events} and scaled by a factor $n/N'_{\rm sim}$. 
Since the bulk of neutrinos considered ($\sim90 \%$) are below $10^4\: \si\GeV$ (see~\cref{fig:edist} (top)), we ignore the neutrino attenuation through Earth since it is known to be small at these energies~\cite{Vincent:2017svp}.

\subsection{Atmospheric neutrino fluxes used to estimate background}

The atmospheric flux $\dot{\phi}_{\rm bkg}$, used to estimate starting and throughgoing background, is obtained using the \texttt{NuFlux}~\cite{NuFluxRepository} interface, which interpolates the neutrino flux computed by \texttt{MCEq}~\cite{Fedynitch:2022vty} assuming the Hillas-Gaisser H3a cosmic ray model~\cite{Gaisser:2013bla}, and the \texttt{Sybill2.3c} hadronic interaction model~\cite{Riehn:2017mfm}.
The solid angle required to calculate the atmospheric flux is based on the conservative detector angular resolutions estimated in Ref.~\cite{Kopp:2007ai}, namely $17 \degree$ for starting events and $5 \degree$ for throughgoing events, integrated for the same time interval as the corresponding SN.


\end{document}